\documentclass[aps,prb,twocolumn,showpacs,10pt]{revtex4-1}
\usepackage{amsmath}
\usepackage{graphicx}
\usepackage{subfigure}
\begin{document}
\title{Magnetic Vortex Guide}

\author{H. Y. Yuan$^{1,2}$ and X. R. Wang$^{1,2,}$}
\email{[Corresponding author:]phxwan@ust.hk}
\affiliation{$^{1}$Physics Department, The Hong Kong University of
Science and Technology, Clear Water Bay, Kowloon, Hong Kong}
\affiliation{$^{2}$HKUST Shenzhen Research Institute, Shenzhen 518057, China}

\begin{abstract}
A concept of magnetic vortex guide is proposed and numerically studied.
Similar to the waveguides of electromagnetic waves, a magnetic vortex guide
allows a vortex domain wall to move along a nanostrip without annihilation
at the strip edges. It is shown by micromagnetic simulations that a magnetic
nanostrip of a properly designed superlattice structure or bilayered
structure can serve as vortex guides.
\end{abstract}

\pacs{75.78.-n, 75.60.Ch, 75.78.Cd, 85.75.-d}
\maketitle

\section{Introduction}
Magnetic vortex, characterized by the in-plane curling magnetization
and out-of-plane magnetization at the core, has attracted much attention
because of its promising applications in high-density storage devices
\cite{Shinjo2000,Wach2002,Choe2004,Waeyen2006,Yu2008,Kim2010,Streubel2012}.
A vortex can attach itself to a magnetic domain wall, forming a vortex
domain wall (VDW). A VDW in a magnetic nanowire is a localized spin texture
and can move around. Thus, a VDW can be a mobile information carrier when
the information is encoded in the vortex core. Although a vortex, as a
topological object, cannot be destroyed or created by itself in sample bulk
unless another vortex with opposite topological number is involved,
it could be annihilated or created at sample surfaces. On the other hand, a
vortex tends to move sideways toward sample surfaces due to gyroscopic effect.
One challenge is how to keep a vortex away from sample surface
when it travels in a nanostrip wire. Here, we propose a concept of magnetic
vortex guide to meet this challenge. Similar to the optical waveguide that
guides electromagnetic wave, the magnetic vortex guide can confine a vortex
to move within a designed region and avoids possible loss of vortices
and the information carried by them.

This paper is organized as follows. The theoretical analysis of vortex
dynamics and vortex velocity is presented in Sec. II. The numerical
realization and discussion of the vortex guide are illustrated in Sec. III
followed by conclusions.

\section{Theoretical analysis}

A VDW in a nanowire can be driven by a static magnetic field via energy
dissipation \cite{Wang2009,Wang20092}, by an electric current via spin transfer
torque (STT) \cite{Berger1984,Slonczewski1996, Zhang2004,Thiaville2005}, or by
a microwave via both energy exchange and angular momentum exchange \cite{yan}.
In this article, we consider only current-driven VDW motion because of both
its fundamental interest and the technological importance.
Generally speaking, STT has both an adiabatic component and a non-adiabatic
component although the adiabatic one is normally much bigger than the other.
The motion of a VDW in a magnetic strip wire is mainly dragged by the motion
of its vortex core when an electric current is passing through the wire.
STT drives a vortex core moving in both transverse and longitudinal directions.
The transverse motion of the vortex core is detrimental for the survival of a
VDW because the vortex is destroyed once its core collides with the wire edge.
The information coded in the vortex will be lost upon this annihilation.
Thus, it is important to keep the vortex core
away from the wire edges if VDWs are used as mobile information carriers.

In order to understand the working principle of the vortex guides proposed
below, we explore analytically the vortex dynamics. The system we consider
is a very thin magnetic film, whose length, width and thickness are respectively
along the $x$, $y$ and $z$-axis of the Cartesian coordinate system.
Generally, the vortex dynamics is governed by the Thiele equation
\cite{Thiele1973, Huber1982, Thiaville2005},

\begin{equation*}
\mathbf{F}+\mathbf{G}\times(\mathbf{v-u})+\mathbf{D}\cdot(\alpha\mathbf{v}
-\beta \mathbf{u})=0,
\end{equation*}
where $\mathbf{F}$ is the external force related to magnetic field that is
zero in our case, $\mathbf{G}$ is gyrovector that is zero for a transverse
wall and $\mathbf{G}=-2\pi qplM_s/\gamma\mathbf{\hat{z}}$ for a 2D vortex
wall, where $q$ is the winding number (+1 for a vortex and -1 for an
antivortex), $p$ is the vortex polarity ($\pm 1$ for core polarization in
$\pm z$ direction), $l$ is the thickness of nanowire and $\gamma$ is
gyromagnetic ratio. $\mathbf{D}$ is dissipation dyadic, whose none zero
elements  for a vortex wall are $D_{xx}=D_{yy}=-2M_s Wl/(\gamma \Delta)$
\cite{Huber1982}, where $W$ is nanowire width and $\Delta$ is the Thiele's
domain wall (DW) width \cite{Thiele1973}. $u=jP\mu_B/(eM_s)$ is proportional to the
current density $j$ and is along the current direction, where $P$, $\mu_B$
and $e$ are the current polarization, the Bohr magneton and the electron
charge, respectively. For permalloy, $u=100$ m/s corresponds to $j=1.38
\times 10^{12}$ $\mathrm{A/m^2}$ ($P=1$). In our consideration,
electrons flow in the $+x$-direction. $\alpha$ and $\beta$ are Gilbert 
damping constant and non-adiabatic strength, respectively. 
Both $\alpha$ and $\beta$ are dimensionless parameters
\cite{Zhang2004,Thiaville2005}.
$\mathbf{v}$ is the DW velocity. One can recast the equation in terms 
of $\mathbf{v}$ as a function of $\mathbf{u}$
\begin{equation*}
\begin{aligned}
&\alpha v_x - \frac{\pi q p \Delta}{W} v_y = \beta u, \\
&v_x +  \frac{\alpha W}{\pi q p \Delta} v_y = u. \\
\end{aligned}
\end{equation*}
Solving these two equations simultaneously gives
\begin{equation*}
\begin{aligned}
&v_x = \frac{u}{1+\alpha^2W^2/(\pi^2 \Delta^2)}\left(1
-\frac{\beta}{\alpha}\right ) + \frac {\beta u}{\alpha},\\
& v_y = \frac{1}{1+\alpha^2W^2/(\pi^2 \Delta^2)} \frac{W}
{\pi q p \Delta} (\alpha - \beta) u.
\end{aligned}
\end{equation*}

As the nanowire width is on the same order of DW width \cite{Nakatani2005}
and the damping constant $\alpha$ is on the order of $10^{-2}$, which
is much smaller than 1, the velocities can be expanded to the leading
order in $\alpha$ and $\beta$ as
\begin{equation}
\begin{aligned}
&v_x \approx u - \frac{\alpha (\alpha - \beta) W^2 u}{\pi^2 \Delta^2},\\
\end{aligned}
\end{equation}
\begin{equation}
\begin{aligned}
& v_y \approx \frac{W}{\pi q p \Delta} (\alpha - \beta) u.
\end{aligned}
\label{vy}
\end{equation}
The formula (\ref{vy}) shows that the direction of the transverse motion of a
vortex core depends sensitively on $\alpha-\beta$.
The transverse velocity is always non-zero if $\alpha \neq \beta $
\cite{Thiaville2005, Beach2006,Hayashi2007,Meier2007}.
For a vortex with a given topological number $pq$, say -1,
its transverse motion is opposite for $\alpha < \beta$ as shown in
Fig. 1a and $\alpha > \beta$ shown in Fig. 1b.

Based on above discussion of the transverse motion of a VDW core, one
natural proposal is to use a superlattice wire, as schematically
illustrated in Fig. 1c, to control the transverse motion of the vortex.
The superlattice is made of two basic blocks A and B with $\alpha-\beta$
taking opposite signs, say $\alpha-\beta<0$ in A and $\alpha-\beta>0$ in B.
A VDW, say with $pq=-1$, shall move upward in A and downward in B.
By carefully designing the lengths of blocks A and B, it is possible to
compensate the transverse motion of a VDW after a complete period so that
the VDW travels longitudinally on average along the superlattice wire
without annihilation at the wire edges. Here it's assumed that the average 
velocity is the mean value of VDW velocities in blocks A and B. 
the average VDW velocity is approximately equal to

\begin{equation}
\begin{aligned}
\bar{v}_x = u - \frac{W^2 u}{2\pi^2 \Delta^2} (\alpha_A^2+\alpha_B^2 - \beta (\alpha_A +\alpha_B))
\end{aligned}
\label{barv}
\end{equation}
where $\alpha_A$ and $\alpha_B$ are respectively the damping constant 
in blocks A and B. It is worth mentioning that
such a superlattice wire is possible to realize with today's technology.
A recent study \cite{Reidy2003} demonstrated that $\alpha$ of permalloy
can increase by four times through a dilute impurity doping of
lanthanides (Sm, Dy, and Ho).


\section{Results and discussions}
To demonstrate the feasibility of such a magnetic vortex guide,
we consider a special superlattice in which $\alpha - \beta$ has
almost the same magnitude but opposite sign in the two unit blocks.
The lengths of block A and block B are 1000 nm so that
the vortex core moves near the medial axis of the superlattice.
In our micromagnetic simulations by using {\footnotesize MUMAX} package
\cite{mumax}, a nanowire of 520 nm wide and 8 nm thick is used.
A moving simulation window of 2000 nm long is always entered at
the vortex core. To mimic permalloy wires, we use the exchange
constant $A = 1.3 \times 10^{-11}$ J/m, saturation magnetization
$M_s = 8 \times 10^5$ A/m, and zero crystalline anisotropy.
The mesh size is $4 \times 4 \times 4$ $\mathrm{nm^3}$.
The non-adiabatic STT coefficient $\beta=0.015$ is used.
The damping coefficients in blocks A and B are
0.01 and 0.0215, respectively. 

Figure 1d is the time-dependence of the vortex core position 
($x_c, y_c$) under a current density of $6.9 \times 10^{12} \mathrm{A/m^2}$, 
corresponding to $u =500$ m/s. ($x_c, y_c$) is the coordinate of the core 
spin with the minimum $z-$component of magnetization.
The initial vortex wall has $q=+1$ and $p=-1$.
After the current is switched on at $t = 0$, the VDW starts to
propagate and enters the guide around $t=1$ ns, indicated by the blue line.
The black and red lines in Fig. 1d are respectively $y_c(t)$ and $x_c(t)$.
The vortex core moves upward by 16 nm for about 2.2 ns in block A then
downward by 16 nm for about 2.2 ns in block B alternatively, as expected.
The linear behavior of $x_c(t)$ shows that the longitudinal velocity of
the vortex core is $v_x=475$ m/s, very close to Thiele equation result.
Figure 1e is the current density dependence of VDW velocity and it agrees 
well with the theoretical prediction (\ref{barv}) shown by the red line.

\begin{figure}
\centering
\includegraphics[width=0.45\textwidth]{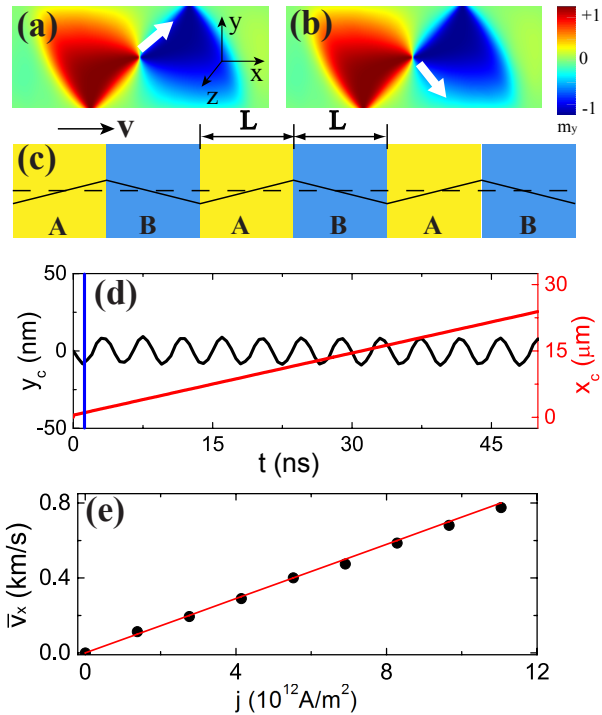}\\
\caption{ (color online)Sketches of moving direction (black arrows) of
a VDW of $q = 1$ and $p=-1$ in a nanowire with $\alpha < \beta$
(\textbf{a}) and $\alpha > \beta$ (\textbf{b}). The white arrows denote
the magnetization directions. The x-, y-, and the z-axis are respectively
along the length direction, the width direction and the depth direction.
The color is coded by the value of $m_y$ as indicated by the right bar.
(\textbf{c}) The superlattice structure of a magnetic vortex guide.
Block A: $\alpha<\beta$. Block B: $\alpha>\beta$.
The zig-zag line sketches the trajectory of a vortex core ($qp < 0$) in
the vortex guide. (\textbf{d}) Vortex core position ($x_c, y_c$) as the
functions of time $t$ under a current density of $6.9 \times 10^{12}$
$\mathrm{A/m^2}$, corresponding to $u =500$ m/s.
The black and red lines are respectively for $x_c$ and $y_c$. The linear
behavior of $x_c$ shows a  longitudinal velocity of 475 m/s. The blue line
corresponds to the time when VDW enters vortex guide.
(\textbf{e}) Current density dependence of vortex velocity. Straight line
is theoretic prediction.}
\label{fig1}
\end{figure}

Obviously, a perfect superlattice VDW guide requires a complete
compensation of transverse motion of a VDW inside blocks A and B.
This is not easy to achieve because the motion of a VDW is sensitive to
many material parameters including both $\alpha$ and $\beta$. Any
fluctuation or uncertainty in these parameters shall lead to an overall
shift in the transverse direction, no matter how small it might be.
After a long time and/or a long travel distance, the vortex might
eventually collide with the wire edges and annihilate there.
Furthermore, our current understanding of current-driven DW motion
does not give us the power to predict the exact compensatory length
for blocks A and B for given sets of material parameters. One can,
at most, use micromagnetic simulations to provide certain guidance
if one wants to build a long VDW guide in reality. Thus, it shall
be ideal if one can make a self-focusing guide for VDWs like an
optical fiber for light. Interestingly, a bilayer wire as shown in
Fig. 2a can confine a VDW with $qp=-1$ near the interface if the
top layer (block B) is made of material with $\alpha > \beta$ and
the bottom layer (block A) is a material with $\alpha < \beta$.
According to above formula for the transverse motion of a VDW, the
VDW moves downward in layer B and upward in layer A.
As a result, the vortex core shall move around the interface during
its lateral propagation along the interface of the layered structures.
To demonstrate such a bilayer wire is capable of confining the core of a
VDW of $qp=-1$ near the interface of the bilayer, we consider a bilayer
wire with $\beta=0.015$ in both layers, and $\alpha=0.01$ in layer A and
$\alpha=0.0215$ in layer B. Both layer A and layer B have the same width of
260 nm and the same thickness of 8 nm. Figure 2b shows $x_c(t)$ and $y_c(t)$ of
a VDW under an electric current density of $6.9 \times 10^{12}$ $\mathrm{A/m
^2}$, corresponding to $u =500$ m/s which is switched on at $t=0$.
An initially created VDW enters the vortex guide around 1 ns.
According to Fig. 2b, $y_c \simeq 0$, the VDW is confined around interface.
The longitudinal motion of the VDW is at a constant speed (linear curve for
$x_c(t)$) of $v_x\simeq 474$ m/s. The transverse motion of the VDW is
actually oscillatory with a very small amplitude (0.4 nm).
This oscillation feature is not captured by the Thiele equation.
The reason may be that a vortex is an extensive and deformable object,
instead of a rigid-body object that can be approximated as a point.
Thus, when its core passes through the interface, the vortex cannot adjust
its structure fast enough so that the force acting on the vortex does not
change its sign. It shall be a challenging problem to generalize the Thiele
equation such that the oscillatory motion of the vortex core near the
interface of the bilayer wire can be described.
Figure 2c is the current density dependence of VDW velocity and it's well
described by the theoretical prediction shown by the red line.

\begin{figure}
\centering
\includegraphics[width=0.45\textwidth]{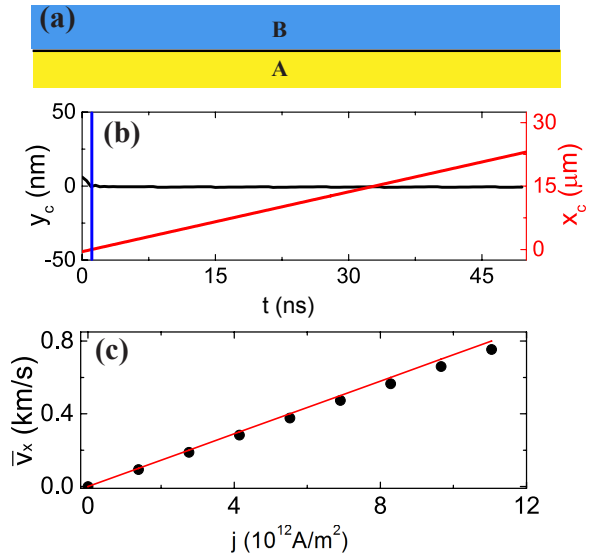}\\
\caption{(color online)
(\textbf{a}) The structure of second type of
magnetic vortex guide. Block A: $\alpha < \beta$. Block B: $\alpha > \beta$.
(\textbf{b}) Vortex core position ($x_c, y_c$) as the functions of
time $t$ under a current density of $6.9 \times 10^{12}$ $\mathrm{A/m^2}$,
corresponding to $u =500$ m/s. The red and black lines are respectively
for $x_c$ and $y_c$. The linear behavior of $x_c$ indicates that the
longitudinal motion of the VDW is at a constant speed of 474 m/s.
(\textbf{c}) Current density dependence of VDW velocity.}
\label{fig2}
\end{figure}

The vortex guide discussed above is only for $qp=-1$ ($q$ is the
winding number and $p$ is the vortex polarity). The vortex guide for
$qp=1$ can be trivially obtained by exchanging layer A with layer B.
Although the results presented here are for strip wires of 520 nm wide and
8 nm thick, in which a vortex wall is stable \cite{McMichael1997, Nakatani2005}, they
are also similar for wires whose widths is so small that a transverse
wall is the most stable DW but a VDW is still a metastable DW.
In order to generate a metastable VDW, a notch assistant process can be used.
It is known that depinning of a transverse wall at a notch starts from
the transformation of a transverse wall into a vortex wall \cite{Yuan2014}.
It is also known that for an extremely small wire, a VDW may not be able
to nucleate. For permalloy, this limit is 24 nm. Thus 24 nm is
the smallest width of a vortex guide if it is made of permalloy.

\section{Conclusions}
In conclusion, the concept of magnetic vortex guide is raised.
Two vortex guide structures are proposed and numerically studied.
One is a superlattice strip whose two unit blocks have opposite
signs of $\alpha-\beta$. Such a guide requires an exact compensation
of the transverse motion of a VDW in blocks A and B, and high
demanding in the accurate design of two basic blocks is needed.
The second type of vortex guide is a simple bilayer strip in which
$\alpha-\beta$ has opposite signs in two layers.
The possible strip width of a magnetic vortex guide is determined
by the minimal strip width below which a VDW cannot exist.

\section*{Acknowledgments}

This work was supported by China NSFC
grant (11374249) and Hong Kong RGC grant (605413).
H.Y.Y. acknowledges the support of Hong Kong PhD Fellowship.


\begin{thebibliography}{}
\bibitem{Shinjo2000} T. Shinjo, T. Okuno, R. Hassdorf, K. Shigeto,
and T. Ono, Science \textbf{289}, 930 (2000).

\bibitem{Wach2002} A. Wachowiak, J. Wiebe, M. Bode, O. Pietzsch,
M. Morgenstern, R. Wiesendanger, Science \textbf{298}, 577 (2002).

\bibitem{Choe2004} S. -B. Choe, Y. Acremann, A. Scholl, A. Bauer,
A. Doran, J. St\"{o}hr, and H. A. Padmore, Science
 \textbf{304}, 420 (2004).

\bibitem{Waeyen2006} B. Van Waeyenberge, A. Puzic, H. Stoll, K. W. Chou,
T. Tyliszczak, R. Hertel, M. F\"{a}hnle, H. Br\"{u}ckl, K. Rott,
G. Reiss, I. Neudecker, D. Weiss, C. H. Back, and G. Sch\"{u}tz,
Nature (London) \textbf{444}, 461 (2006).

\bibitem{Yu2008} K. Y. Guslienko, K. -S. Lee, S. -K. Kim,
Phys. Rev. Lett. \textbf{100}, 027203 (2008).

\bibitem{Kim2010} D. -H. Kim, E. A. Rozhkova, I. V. Ulasov,
S. D. Bader, T. Rajh, M. S. Lesniak, and V. Novosad,
Nat. Mater. \textbf{9}, 165 (2010).

\bibitem{Streubel2012} R. Streubel, D. Makarov, F. Kronast,
V. Kravchuk, M. Albrecht, and O. G. Schmidt,
Phys. Rev. B \textbf{85}, 174429 (2012).

\bibitem{Wang2009} X. R. Wang, P. Yan, J. Lu, and C. He, 
Ann. Phys. (N.Y.) \textbf{324}, 1815 (2009)

\bibitem{Wang20092} X. R. Wang, P. Yan, and J. Lu,
Europhys. Lett. \textbf{86}, 67001 (2009).

\bibitem{Berger1984} L. Berger, J. Appl. Phys. \textbf{55}, 1954 (1984).

\bibitem{Slonczewski1996} J. C. Slonczewski,
J. Magn. Magn. Mater. \textbf{159}, L1 (1996).

\bibitem {Zhang2004} S. Zhang and Z. Li,
Phys. Rev. Lett. \textbf{93}, 127204 (2004).

\bibitem{Thiaville2005} A. Thiaville, Y. Nakatani, J. Miltat, and
Y. Suzuki, Europhys. Lett. \textbf{69}, 990 (2005).

\bibitem{yan} P. Yan and X. R. Wang,
Phys. Rev. B \textbf{80}, 214426 (2009).

\bibitem{Thiele1973} A. A. Thiele, Phys. Rev. Lett. \textbf{30}, 230 (1973).

\bibitem{Huber1982} D. L. Huber, Phys. Rev. B \textbf{26}, 3758 (1982).

\bibitem{Nakatani2005} Y. Nakatani, A. Thiaville, and J. Miltat,
J. Magn. Magn. Mater. \textbf{290-291}, 750 (2005).

\bibitem{Beach2006} G. S. D. Beach, C. Knutson, C. Nistor, M. Tsoi,
and J. L. Erskine, Phys. Rev. Lett. \textbf{97}, 057203 (2006).

\bibitem{Hayashi2007} M. Hayashi, L. Thomas, C. Rettner, R. Moriya,
and S. S. P. Parkin, Nat. Phys. \textbf{3}, 21 (2007).

\bibitem{Meier2007} G. Meier, M. Bolte, R. Eiselt, B. Kr\"{u}ger,
D. -H. Kim, and P. Fischer, 
Phys. Rev. Lett. \textbf{98}, 187202 (2007).

\bibitem{Reidy2003} S. G. Reidy, L. Cheng, and W. E. Bailey,
Appl. Phys. Lett. \textbf{82}, 1254 (2003).

\bibitem{mumax} A. Vansteenkiste, J. Leliaert, M. Dvornik, M. Helsen,
F. Garcia-Sanchez, and B. Van Waeyenberge,
 AIP Adbances \textbf{4}, 107133 (2014).

\bibitem{McMichael1997} R. D. McMichael and M. J. Donahue,
IEEE. Trans. Magn. \textbf{33}, 4167 (1997).

\bibitem{Yuan2014} H. Y. Yuan and X. R. Wang,
arXiv:1407.4559 [cond-mat.mes-hall].

\end{thebibliography}
\end{document}